\documentstyle[11pt,paspconf]{article}

\newcommand{\lya}{\mbox{${\rm Ly}\alpha$}}
\newcommand{\etal}{et al.}

\begin{document}

\title{Photometric/Spectroscopic Redshift Identification of Faint Galaxies in STIS Slitless Spectroscopy Observations}
\author{Hsiao-Wen Chen\altaffilmark{1}, Kenneth M. Lanzetta\altaffilmark{1}, 
and Sebastian Pascarelle\altaffilmark{1}}
\affil{Department of Physics and Astronomy, State University of New York at 
Stony Brook, Stony Brook, NY 11794-3800}

\begin{abstract}

  We present a new spectrum extraction technique which employs optimal weights 
for the spectral extraction, deblends the overlapping spectra, determines the 
precise sky background, and takes into account correlations between errors 
correctly for STIS slitless observations. We obtained roughly 250 optimally 
extracted spectra in a deep STIS field as well as self-confirming redshift 
measurements for these objects, including a galaxy at $z=6.68$. In addition, 
we identified five isolated emission-line objects in the dispersed image that 
were not accounted for by objects detected in the direct image. Assuming that 
these are \lya\ emission line galaxies at high redshifts and adopting a simple 
$\delta$-function galaxy luminosity function for these objects, we find, based 
on a likelihood analysis, that \lya\ emission line galaxies at 
$\langle z\rangle\approx 4.69$ may contribute at most 15\% of the star 
formation rate density measured in Lyman break galaxies at $z\approx 4$.

\end{abstract}


\section{Introduction}

  Observations of distant galaxies have advanced rapidly as a result of the 
Keck telescope, which because of its large collecting area is sensitive to 
faint objects.  But despite extensive searches, only four galaxies of redshift
$z > 5$ have as yet been identified spectroscopically using the Keck telescope
(Dey et al.\ 1998; Hu, McMahon, \& Cowie 1998; Weymann et al.\ 1998; Spinrad et 
al.\ 1998).  It has proven extremely difficult to identify high-redshift 
galaxies solely based on ground-based spectroscopy, because (1) galaxies lack 
prominent narrow-band features at rest-frame ultraviolet wavelengths (which are
redshifted to observed-frame optical or infrared wavelengths) and (2) 
background sky light is the dominant source of noise at near-infrared 
wavelengths.  

  We have sought to identify distant galaxies in very deep slitless 
spectroscopy observations acquired by STIS on board HST, by combining a new 
spectrum extraction technique with photometric and spectroscopic analysis 
techniques. Our analysis was designed to identify redshifts of galaxies in the 
slitless data by means of broad-band photometric techniques and to test these 
redshifts by identifying narrow-band emission, absorption, and continuum 
features in the same spectra. As a result of this analysis, we obtained 
optimally extracted spectra and self-confirming redshift measurements for 
roughly 250 object, including a galaxy at $z=6.68$, and 5 isolated 
emission-line objects. 

\section{Data}

    The very deep observations, obtained by HST using STIS toward a region of 
sky flanking the Hubble Deep Field, consisted of pairs of images:  a direct 
image taken using no filter and a dispersed image taken using the G750L 
grating.  Additional observations consisted of only a direct image. The 
integration time of the direct images totaled 4.5 h over 82 exposures, and the 
integration time of the dispersed images totaled 13.5 h over 60 exposures.
We summed the direct and dispersed images using conventional image processing
techniques. The summed image covers a sky area of $51 \times 51$ arcsec$^2$.
The spatial resolution of the summed direct and dispersed images is ${\rm FWHM}
\approx 0.08$ arcsec, the $1 \sigma$ detection threshold of the summed direct 
image is $\approx 26.2$ mag arcsec$^{-2}$, and the $1 \sigma$ detection 
threshold of the summed dispersed image at $\lambda \approx 9800$ \AA\ is 
$\approx 5.5 \times 10^{-18}$ erg s$^{-1}$ cm$^{-2}$ \AA$^{-1}$ arcsec$^{-1}$. 

\section{Optimal Extraction of Slitless Spectra}

  The spectrum extraction of the dispersed image is made especially difficult 
because the image of the field is covered by light of faint galaxies, which
when dispersed overlaps in the spatial direction and is blurred in the spectral
direction.  To spatially deblend and spectrally deconvolve the spectra, we used
the summed direct image to determine not only the exact object locations but 
also the exact two-dimensional spatial profiles of the spectra on the summed 
dispersed image. The spatial profiles of these objects are crucial because (1) 
they provide the ``weights'' needed to optimally extract the spectra, (2) they 
provide the models needed to deblend the overlapping spectra and determine the 
background sky level, and (3) they provide the spectral templates needed to 
optimally deconvolve the spectral blurring of extended objects. 

  First,  we identified objects in the summed direct image, using the 
SExtractor program of Bertin \& Arnouts (1996).  Roughly 250 objects were 
identified in the summed direct image.  Next, we modeled each pixel $(i,j)$ of 
the summed dispersed image as a linear sum of contributions from (1) relevant 
portions of all overlapping neighboring objects and (2) background sky:
\begin{equation}
{\cal F}_{i,j}=\sum_{k}S^k_{i-i_k-\Delta+1}\sum_{i^{''}_k}f^k_{i_k-i^{''}_k,j} + B_{i,j},
\end{equation}
where $S^k_l$ are spectral elements of the $k$th object, $i_k$ is the object 
position along the spectral direction in the direct image, $\Delta$ is the 
constant offset between object position and the starting pixel of the spectrum 
along the spectral direction, $f$ is the object profile measured in the 
direct image, and $B$ is the model sky (for which we found that a fourth order 
polynomial is necessary and sufficient).

  We treated the problem as a $\chi^2$ minimization problem, where the data 
were the pixel values of the summed dispersed image, the model was a linear sum
of appropriate elements of the spatial profiles, and the parameters of the 
model were values of the spectral pixels.  The $\chi^2$ is written
\begin{equation}
\chi^2=\sum_{i,j}\frac{({\cal F}_{i,j}-
\tilde{\cal F}_{i,j})^2}{\sigma_{i,j}^2},
\end{equation}
where $\tilde{\cal F}_{i,j}$ is the flux measurement and $\sigma_{i,j}$ is the 
1-$\sigma$ error at pixel $(i,j)$.
  
  Finally, we minimized $\chi^2$ between the model and the data with respect 
to the model parameters and obtained estimates of all model parameters
simultaneously to form one-dimensional spectra. The model parameters 
included roughly 250,000 ``object'' parameters (from roughly 1000 spectral 
pixels each of roughly 250 objects) and four thousand more ``sky'' parameters.
As a zeroth order approximation, the spectral elements of individual objects 
are independent of each other, and so we can solve the minimum $\chi^2$ 
separately for individual columns. The $\chi^2$ for column $i$ is
\begin{equation}
\chi_i^2=\sum_j\frac{\left(\sum\limits_{k}S^k_{i-i_k-\Delta+1}\sum\limits_{i^{''}_k}f^k_{i_k-i^{''}_k,j}+\sum\limits_{\alpha=0}^L a_{i}^{\alpha}j^{\alpha} -
\tilde{\cal F}_{i,j}\right)^2}{\sigma_{i,j}^2}.
\end{equation}
Now we only need to solve for approximately 250 spectral elements $S^k_l$ and 
$(L+1)$ sky paremeters $a_i^{\alpha}$ (here $L=4$) that give the minimum 
$\chi^2_i$ each time and repeat the calculation for all columns. Errors of the 
model parameters were obtained by solving the Hessian matrix at the minimum 
$\chi^2_i$. 

  This new spectrum extraction method is superior to other extraction methods 
in many ways.  First, the method employs optimal weights for the spectral 
extraction.  Second, the method deblends the spectra.  Third, the method 
determines the precise sky background.  Because the dispersed image is covered 
by light of extremely faint galaxies over most of its area, the spatial 
templates are needed to find the rare ``clear'' regions of sky background 
between the object spectra. Finally, the method estimates errors correctly by
taking into account the correlation between spectral elements of overlapping
objects.

  As a result of the spectrum extraction technique, we obtain 250 optimally 
extracted spectra for redshift analysis. To determine redshifts of extremely 
faint galaxies in STIS slitless spectra, we first measured photometric 
redshifts by means of a redshift likelihood technique (Fern\'andez-Soto, 
Lanzetta, \& Yahil 1999) and verified these redshifts by identifying 
narrow-band emission, absorption, and continuum features in the same spectra.  
The goal is to obtain self-confirming photometric redshift estimates.

\section{Noise Characteristics}

\begin{figure}[th]
\includegraphics{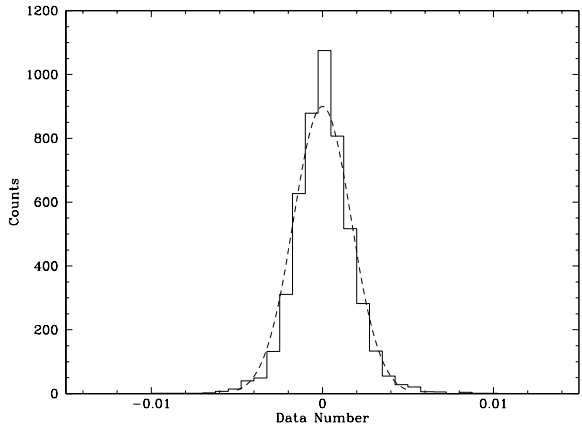}
\vspace{2.25in}
\caption{Histogram of noise in the residual image. The dashed curve indicates 
a best-fit Gaussian profile.}
\end{figure}
    
  The new spectrum extraction technique adopts object profiles determined from 
the direct image as model templates to determine the exact sky and to optimally
extract spectra from the dispersed image. The analysis involves complicated 
smoothing and deblending. It is therefore not unreasonable to expect that the 
noise characteristics may have been highly skewed in the extracted spectra and 
do not follow a normal distribution any more. To proceed with discussions on 
the significance of detections of any spectral features, it is indeed necessary
for us to first understand the noise characteristics of the final products of
the analysis. 

  To examine the noise characteristics, we form a histogram of pixel values
of about 15 rows randomly chosen from the residual image. The histogram is 
shown in Figure 1, in which we also plot the best-fit Gaussian distribution 
with the full width at half maximum of 0.004. The corresponding 1-$\sigma$ 
deviation is 0.0017, which is slightly larger than the 1-$\sigma$ deviation 
measured in the dispersed image (0.0015). The fact that the noise in the
residual image follows a Gaussian distribution with a slightly larger 
1-$\sigma$ deviation has secured our subsequent reference to a normal noise 
distribution for any further statistical analysis. 

\section{Properties of Isolated Emission Lines}

  In addition to the 250 objects observed in the direct image, which included
a galaxy at $z=6.68$, we identified five isolated emission lines in the 
dispersed image that were not accounted for by objects detected in the direct 
image. The $z=6.68$ galaxy is the most distant galaxy that has yet been 
spectroscopically identified. Detailed analysis has been presented elsewhere
(Chen, Lanzetta, \& Pascarelle 1999). Here we focus on the five emission-line
objects. 

\begin{figure}[t]
\includegraphics{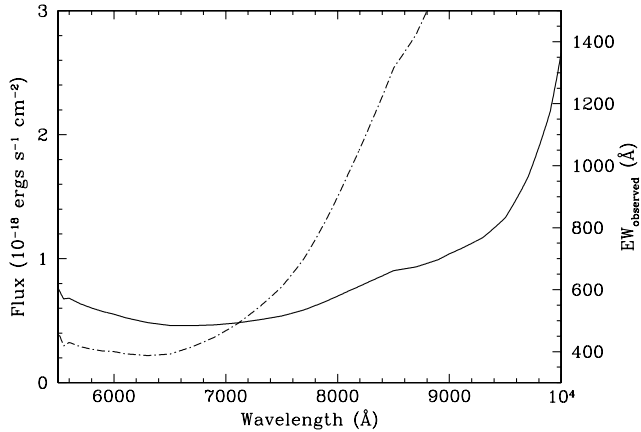}
\vspace{2.25in}
\caption{The solid curve is the 1-$\sigma$ flux threshold of an unresolved 
emission line in the dispersed image versus wavelength. The dot-dashed curve 
shows the 3-$\sigma$ lower limit to the observed equivalent width.}
\end{figure}

  To objectively identify these isolated emission lines, we applied the 
SExtractor program to the smoothed dispersed image, setting the detection 
threshold such that nothing was detected in the negative of the image. Next, we
removed the lines that correspond to objects identified in the direct 
image. Consequently, five isolated emission lines remained in the dispersed 
image. 

  The nature of these isolated emission lines is uncertain because of the 
ambiguities in the wavelength determination. It is impossible to determine the 
wavelengths and therefore to calibrate the fluxes for these lines without first
knowing their positions on the sky. We show in Figure 2 that the 1-$\sigma$ 
flux threshold of an unresolved emission line in the dispersed image may vary 
with wavelength by as much as a factor of four across the entire spectral 
range.  However, given the 1-$\sigma$ single pixel detection threshold of the 
direct image, $\approx 26.2$ mag arcsec$^{-2}$, we can work out the 3-$\sigma$ 
lower limit to the observed equivalent width (EW) as a function of wavelength 
for these isolated emission lines. The dot-dashed curve in Figure 2 indicates 
the 3-$\sigma$ lower limit to the observed EW of an unresolved line detected 
at a 3-$\sigma$ significance level in the dispersed image versus wavelength. 
The actual observed EW limit as a function of wavelength for a particular line 
can be obtained by scaling the dot-dashed curve to the significance of the line
detection. According to the EW limits shown in Figure 2, the observed five
emission lines are most likely to be high-redshift \lya\ emission lines, 
rather than low-redshift [O\,II] or H$\alpha$ lines.

  The sensitivity curve shown in Figure 2 also indicates that the STIS 
observations are more sensitive to faint emission-line galaxies than existing 
deep narrowband surveys from the ground (e.g. Hu, Cowie, \& McMahon 1998). If
these lines are high-redshift \lya\ emission lines, the statistics drawn from
this observation may place a strong constraint on the number density of 
high-redshift \lya\ emission line galaxies and the star formation rate density 
contributed by this population. However, the study has been made extremely 
difficult because of the ambiguities in the wavelength (and therefore redshift)
determination.

\begin{figure}[t]
\includegraphics{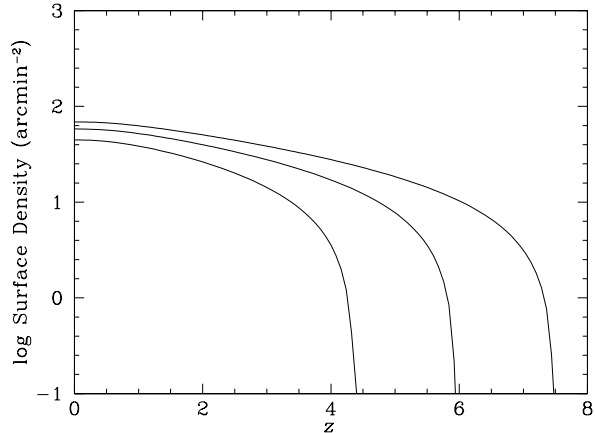}
\vspace{2.25in}
\caption{The predicted surface density of \lya\ emission line galaxies at 
different detection thresholds. The curves, from bottom up, correspond to 
a threshold of 10, 5, and 3 $\times 10^{-18} {\rm ergs}\,{\rm s}^{-1}{\rm 
cm}^{-2}$, respectively.}
\end{figure}

  To solve this problem, we applied a likelihood analysis to the observed 
emission lines, assuming that all five emission lines are high-redshift \lya\ 
lines and that the number density of these lines is well represented by a 
simple $\delta$-function luminosity function,
\begin{equation}
\phi(L/L_*)=\phi_*\delta(L/L_*-1).
\end{equation}
The likelihood analysis returns a best-fit characteristic luminosity $L_*=0.4 
\times 10^{42}\ h^{-2}{\rm ergs}\,{\rm s}^{-1}$. Given the fact that five lines
were detected, we find the number density to be 
$\phi_*=0.04\ h^3{\rm Mpc}^{-3}$. 

  On the basis of the simple $\delta$-function galaxy luminosity function 
determined for the observed \lya\ emission line objects, we estimated the 
statistical properties of the galaxy population. First, we calculated the mean 
redshift of the observed lines and found $\langle z\rangle = 4.69$.  Second, we
estimated the star formation rate density, using the relationship between 
H$\alpha$ luminosity and star formation rate of Madau, Pozzetti, \& Dickinson 
(1998) and assuming case B recombination, where \lya/H$\alpha \approx 10$ 
(Osterbrock 1989). It turns out that the star formation rate density of \lya\ 
emission line galaxies is $0.01\ h\,{\rm M}_{\odot}{\rm yr}^{-1}{\rm Mpc}^{-3}$
for $q_0=0.5$ (without correction for extinction). This is about 15\% of the 
star formation rate density measured by Steidel \etal\ (1999) for the Lyman 
break galaxies at $z>4$, suggesting that galaxies exhibiting the \lya\ 
emission feature represent only a small portion of the whole galaxy population 
at high redshifts. We therefore conclude that high-redshift galaxies are more
efficiently and objectively identified using broad-band photometric rather than
narrow-band imaging/spectroscopic techniques.

  Finally, we calculated the surface density of \lya\ emission line galaxies 
expected in observations of a given sensitivity.  Figure 3 shows that we 
expect to detect no more that five \lya\ emission line galaxies per arcmin$^2$ 
at $z>4$ that are stronger than 
$1\times 10^{-17} {\rm ergs}\,{\rm s}^{-1}{\rm cm}^{-2}$ and no more than 10 
\lya\ emission line galaxies per arcmin$^2$ at $z>5$ that are stronger than 
$5\times 10^{-18} {\rm ergs}\,{\rm s}^{-1}{\rm cm}^{-2}$.

\acknowledgments

  This research was supported by NASA grant NACW-4422 and NSF grant AST-9624216.

\end{document}